\documentclass[twocolumn,aps,prl,nobibnotes,superscriptaddress,amsfonts]{revtex4-1}

\newcounter{firstbib} 

\usepackage{amsmath}
\usepackage{amssymb}
\usepackage{bm}
\usepackage{graphicx}
\usepackage{xfrac}

\newcommand{\mathbfsf}[1]{\ensuremath{\bm{#1}}}

\newcommand{\bv}[1]{\ensuremath{\bm{#1}}}
\newcommand{\dbl}{\ensuremath{ \!\uparrow\! \downarrow \, }}
\newcommand{\spup}{\ensuremath{ \!\!\uparrow }}
\newcommand{\spdn}{\ensuremath{ \!\!\downarrow}}
\newcommand{\ket}[1]{\ensuremath{|{#1}\rangle}}
\newcommand{\expect}[1]{\ensuremath{\langle{#1}\rangle}}
\newcommand{\hhh}{\ensuremath{(\protect\raisebox{0.26em}{\protect\scalebox{0.75}[0.75]{-}}\sfrac{1}{2}\ \,\protect\raisebox{0.26em}{\protect\scalebox{0.75}[0.75]{-}}\sfrac{1}{2}\ \,\sfrac{1}{2})}}

\usepackage{soul,xcolor}

\newcommand{\FigureOne}{
\begin{figure}
\centering\includegraphics[width=\columnwidth]{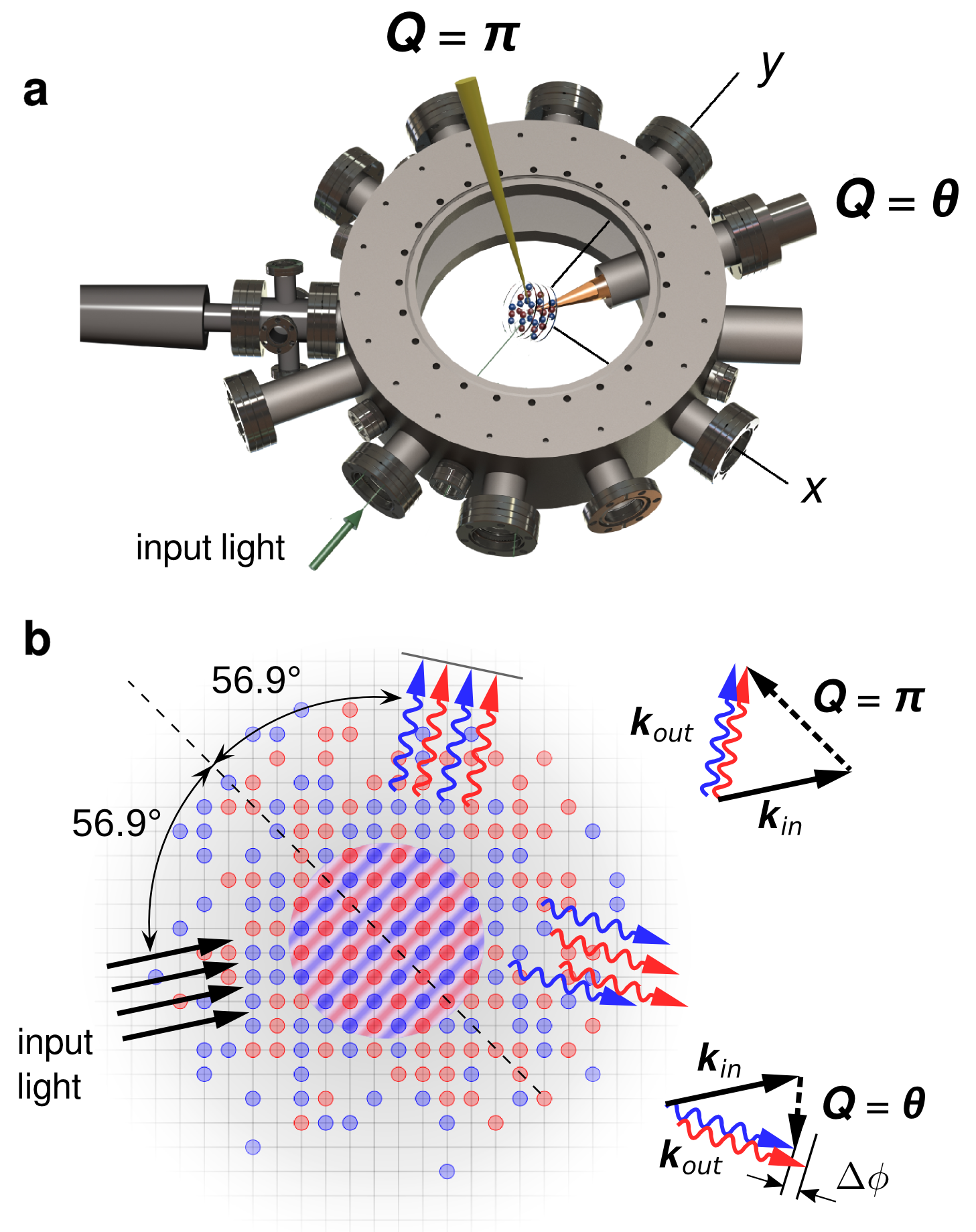}
\caption{\textbf{Schematic depiction of Bragg scattering.} \textbf{a},
Rendering of the experimental setup used for Bragg scattering.  Light is
collected for momentum transfers $\mathbfsf{Q}=\bv{\pi}$ and
$\mathbfsf{Q}=\bv{\theta}$. 
A bias magnetic field, which sets the 
quantisation axis and the interaction strength, points in the $z$ direction.  
The input Bragg beam lies in the $yz$ plane, and its wavevector makes an angle 
of 3$^{\circ}$ with the positive $y$ axis. \textbf{b}, The two spin states are 
denoted by red and blue circles.  AFM order develops at the Mott plateau, shown 
here to be located in the centre, where $n\simeq1$. AFM correlations are 
suppressed outside the central region where $n<1$.  Bragg scattering requires 
the input and output wavevectors, $\mathbfsf{k}_{in}$ and $\mathbfsf{k}_{out}$, 
respectively, to satisfy the Bragg condition 
$\mathbfsf{k}_{out}-\mathbfsf{k}_{in}=\bv{\pi}$.  The red and blue arrows 
denote light scattered from one spin state or the other.  The two spin states 
scatter with opposite phase shift, so that their respective sublattices 
interfere constructively for $\mathbfsf{Q}=\bv{\pi}$.  For a different momentum 
transfer $\mathbfsf{k}_{out}-\mathbfsf{k}_{in}=\bv{\theta}$, scattering is 
relatively insensitive to AFM correlations due to the lack of constructive 
interference between the scattered photons, which have random relative phases 
$\Delta\phi$.}
\end{figure}
}

\newcommand{\FigureTwo}{
\begin{figure}
\centering\includegraphics[width=\columnwidth]{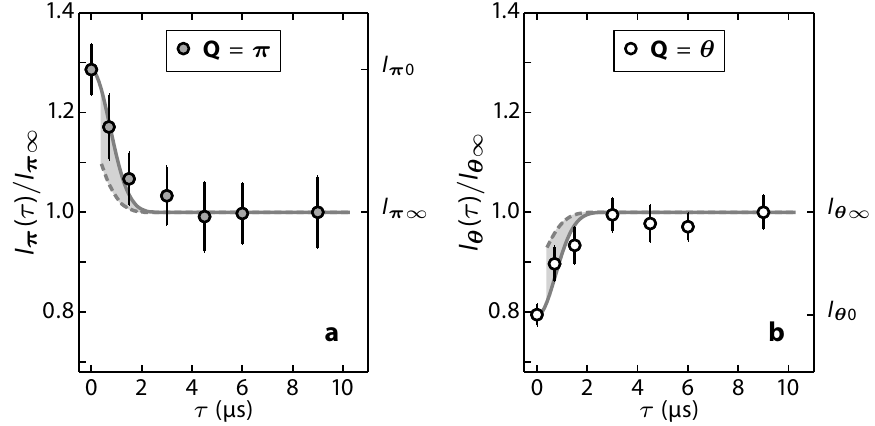} %
\caption{\textbf{Time-of-flight measurement of scattered intensity from a
sample with AFM correlations.}
\textbf{a}, Normalised intensity of Bragg scattered light
($\mathbfsf{Q}=\bv{\pi}$) as a function of time-of-flight $\tau$.  The
\textit{in-situ} ($\tau=0$) scattered intensity is denoted as $I_{\bv{Q}0}$,
while the intensity after sufficiently long $\tau$, corresponding to an
effectively uncorrelated sample, is denoted as $I_{\mathbfsf{Q}\infty}$.
\textbf{b}, For $\mathbfsf{Q}=\bv{\theta}$  the \textit{in-situ}  sample shows
a reduction of scattering, as compared to long $\tau$, due to the presence of
double occupancies and to the presence of AFM spin correlations (see text).
Each data point and error bar is the mean and standard error of the mean
(s.e.m.) of at least 17 measurements of the scattered intensity.  The grey
solid line is the intensity calculated using the value of the Debye-Waller
factor at $\tau$, whereas the dashed grey line uses the average value of the
Debye-Waller factor during the $1.7\,\mu\text{s}$ exposure of the Bragg probe
(see text and Methods).}
\end{figure}
}

\newcommand{\FigureThree}{
\begin{figure}
\centering\includegraphics[width=\columnwidth]{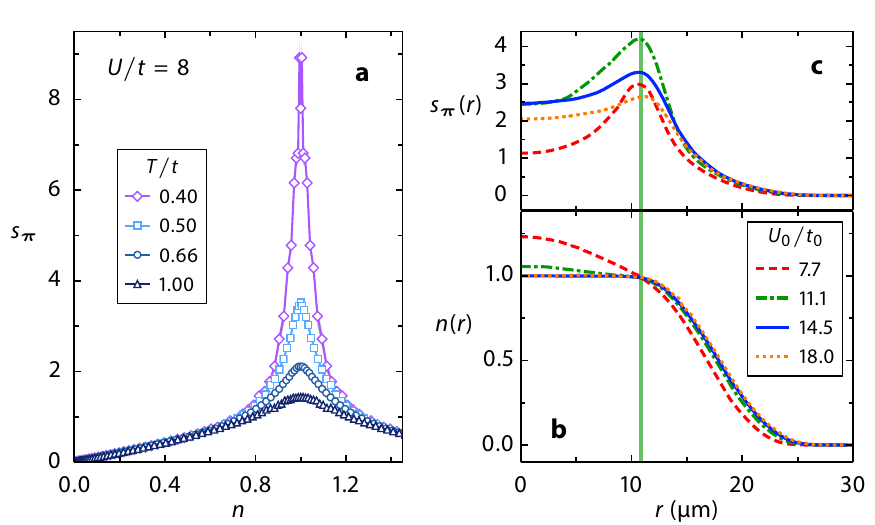}
\caption{\textbf{Numerical calculations} \textbf{a}, Spin structure factor per
lattice site $s_{\bv{\pi}}$ as a function of $n$ in a homogeneous lattice for
several temperatures (see Methods).  $s_{\bv{\pi}}$ is sharply peaked near
$n=1$ and diverges as $T$ approaches $T_{N}$.  \textbf{b}, Density profiles
calculated at $T/t_{0}=0.6$ for different $U_{0}/t_{0}$, using in each case the
value of $N$ that maximises the experimentally measured $S_{\pi}$ (see text and
Extended Data Fig.~2.) \textbf{c}, Profiles of the local spin structure factor
$s_{\bv{\pi}}(r)$, for the same conditions as in \textbf{b}.  The vertical
green line in panels \textbf{b} and \textbf{c} marks the radius at which
$s_{\bv{\pi}}(r)$ is maximised for $U_{0}/t_{0}=11.1$ (see text). }
\end{figure}
}

\newcommand{\FigureFour}{
\begin{figure}
\centering\includegraphics[width=\columnwidth]{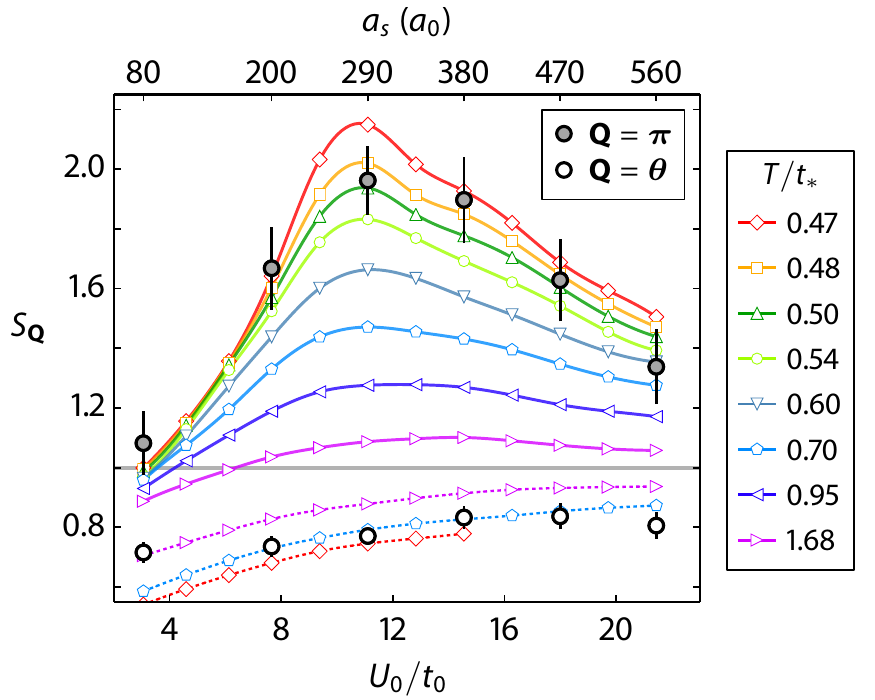}
\caption{\textbf{Spin structure factor}  Measured $S_{\bv{\pi}}$ (filled
circles) and $S_{\bv{\theta}}$ (open circles) at optimised $N$ (see text) for
various $U_{0}/t_{0}$. The values of the $s$-wave scattering length
corresponding to $U_{0}/t_{0}$ for the experimental points are shown along the
top axis.  For each point at least 40 \textit{in-situ} and 40 time-of-flight
measurements of the scattered intensities are used to obtain the spin structure
factor.  Error bars are obtained from the s.e.m. of the scattered intensities;
the raw data is presented in Extended Data Fig.~5.  Numerical calculations of
$S_{\bv{\pi}}$ (open symbols, lines as guide to the eye) and $S_{\bv{\theta}}$
(open symbols, dashed lines as guide to the eye) for various values of
$T/t_{*}$. The numerical calculations for $S_{\bv{\theta}}$ are unreliable for
$T/t_{*}<0.7$ and $U_{0}/t_{0}>15$.  $S_{\bv{\theta}}$ decreases slightly for
weak interactions, where the fraction of double occupancies increases.  }
\end{figure}
}

\newcommand{\ExtFigOne}{
\begin{figure}[H]
\centering
\includegraphics[width=0.65\columnwidth]{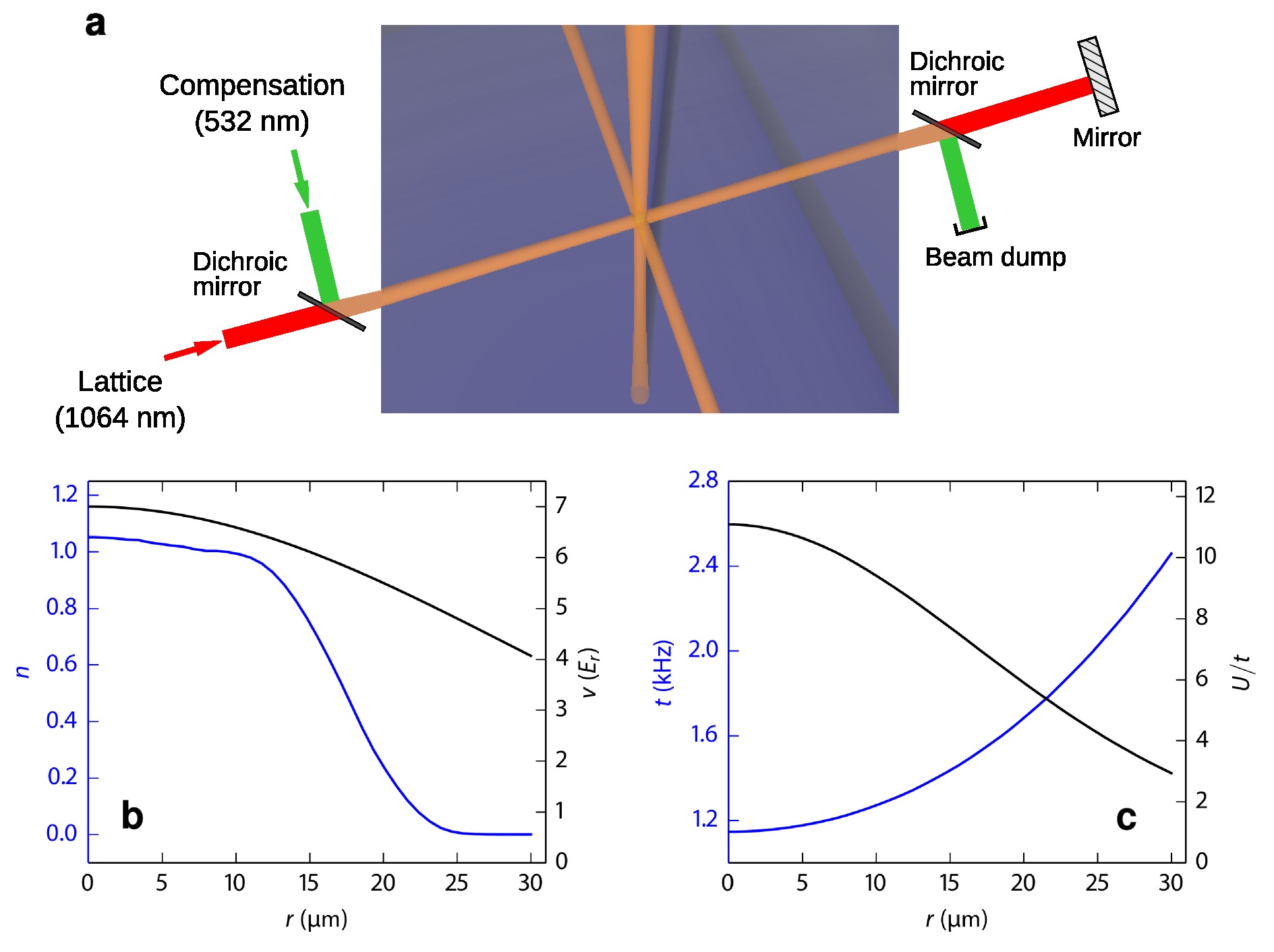}
\caption{\textbf{Compensated optical lattice.} \textbf{a}, Schematic
of the compensated optical lattice setup.  Along each axis, the radial
confinement of the lattice is compensated with a repulsive compensation beam
which copropagates with the lattice beam but is not retroreflected.
\textbf{b},  The local value of the lattice depth $v$ (black line) is shown as
a function of distance from the centre along a body diagonal of the lattice.
Due to the finite extent of the lattice beams, $v$ varies across the density
profile of the cloud (blue line),  which here is calculated for
$U_{0}/t_{0}=11.1$ at $T/t_{0}=0.60$.  \textbf{c},  The inhomogeneity in $v$
results in spatially varying Hubbard parameters $t$ (blue line) and $U/t$
(black line).} 
\end{figure}
}

\newcommand{\ExtFigTwo}{
\begin{figure}[H]
\centering \includegraphics[width=0.45\columnwidth]{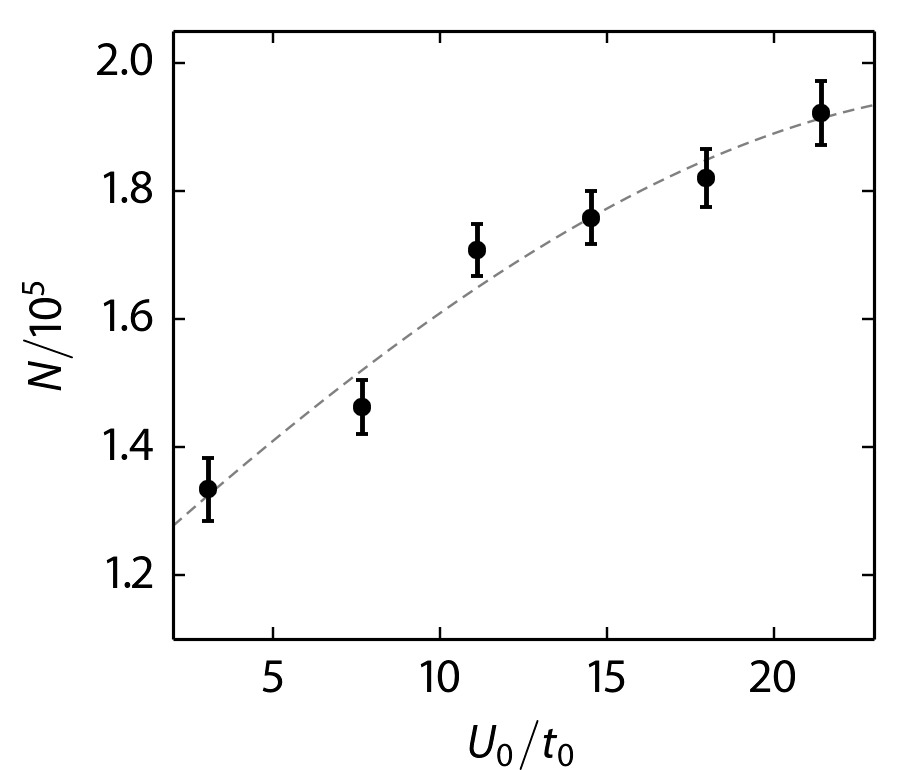}
\caption{\textbf{Atom number for the data in Fig.~4.} Atom number $N$
which maximises $S_{\bv{\pi}}$ as a function of $U_{0}/t_{0}$.  We control $N$
by adjusting the depth of the dimple trap.  Using a linear calibration between
the depth of the dimple trap and the final atom number,  we obtain the value of
$N$ corresponding to the data in Fig.~4.  The error bars correspond to the
s.e.m of the dimple depths used in at least 40 \textit{in-situ}
and 40 time-of-flight realisations of the experiment, corresponding to the data
in Fig.~4.   The line is a third order polynomial fit, which is used to
interpolate the value of $N$ for numerical calculations shown in Fig.~4. }
\end{figure}
}

\newcommand{\ExtFigThree}{
\begin{figure}[H]
\centering \includegraphics[width=0.45\columnwidth]{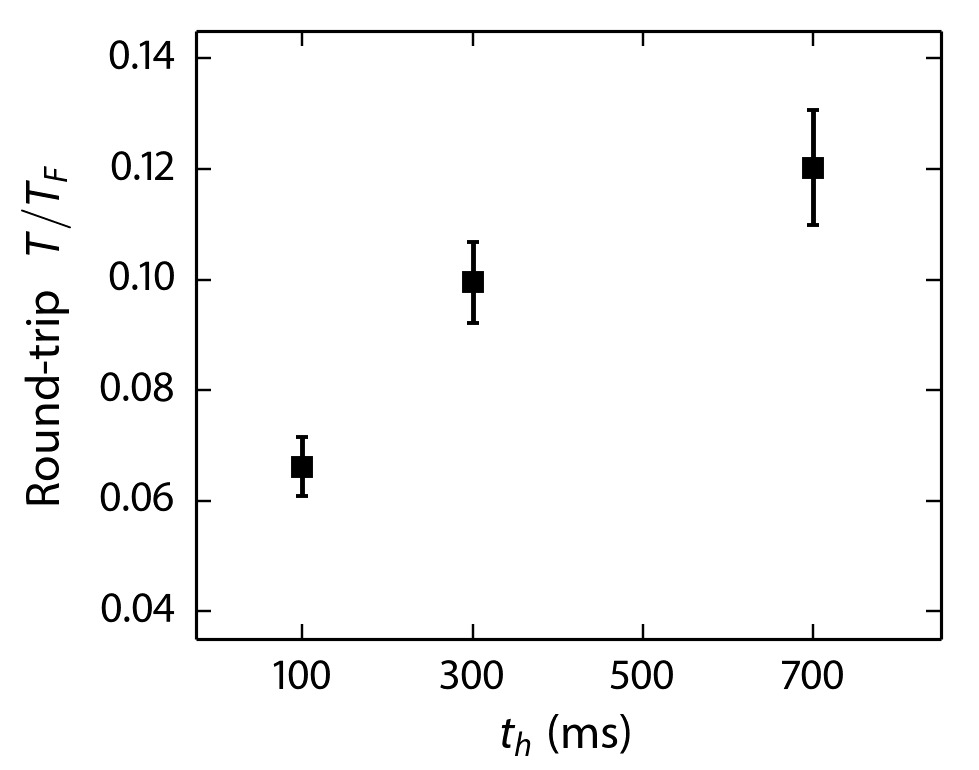}
\caption{\textbf{Round-trip temperature measurements.} Measurement of the
round-trip $T/T_{F}$ vs. hold time $t_{h}$ in a compensated lattice with
$v_{0}=7\,E_{r}$  and $g_{0}=3.7\,E_{r}$.  The duration of the loading ramps is
not included in $t_{h}$.  The scattering length is $326\,a_{0}$,  which
corresponds to $U_{0}/t_{0}=12.5$.  Error bars are the s.e.m. of 6
independent realisations.  The temperature in the dimple trap before loading
into the lattice is $T/T_{F}=0.04\pm0.02$. }
\end{figure}
}
\newcommand{\ExtFigFour}{
\begin{figure}[H]
\centering \includegraphics[width=135mm]{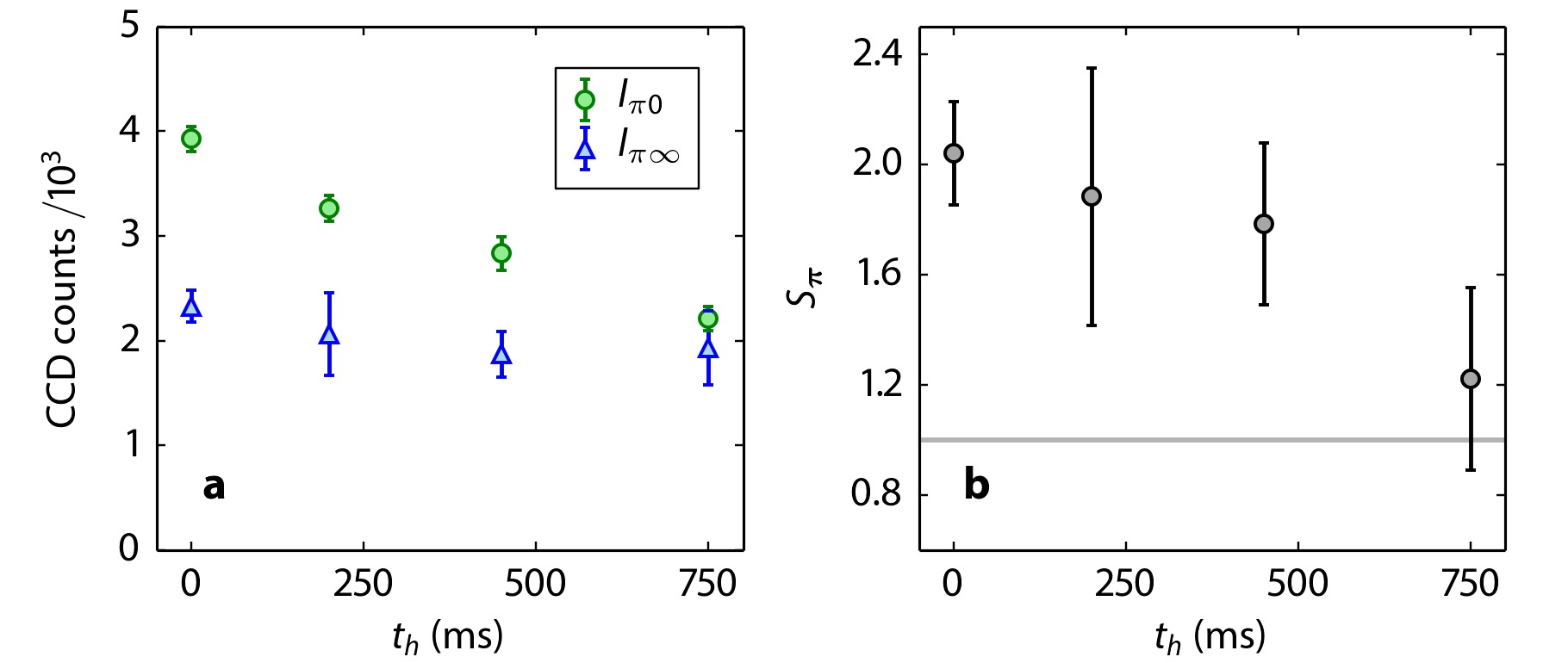} 
\caption{\textbf{Bragg signal decay with hold time.} \textbf{a},
Detected counts vs. $t_{h}$, measured for momentum transfer
$\mathbfsf{Q}=\bv{\pi}$ for an \textit{in-situ} sample ($I_{\bv{\pi}0}$, green
circles) and after decay of the Debye-Waller factor ($I_{\bv{\pi}\infty}$, blue
triangles).  For longer hold times, the Bragg scattered intensity
$I_{\bv{\pi}0}$ decays to match $I_{\bv{\pi}\infty}$,  reflecting the absence
of AFM correlations in a sample at higher $T$.  \textbf{b}, The spin structure
factor corresponding to the scattered intensities shown in \textbf{a}.  For
these measurements the scattering length is $200\,a_{0}$, corresponding to
$U_{0}/t_{0}=7.7$ in a $7\,E_{r}$ deep lattice.  The compensation is
$g_{0}=4.05\,E_{r}$, different from that used for the data in Fig.~4.  The
increased compensation requires a larger atom number to realise an $n\simeq 1$
shell in the cloud.  The atom number used here is $2.6\times10^{5}$ atoms.  The
duration of the Bragg probe is 2.7$\,\mu$s for these data.  Error bars in
\textbf{a} are the s.e.m. of at least 5 measurements for
$I_{\bv{\pi}\infty}$ and at least 10 measurements for $I_{\bv{\pi}0}$.  Error
bars in \textbf{b} are obtained from the s.e.m. of the measured
intensities and equation (2).} 
\end{figure}
}
\newcommand{\ExtFigFive}{
\begin{figure}[H]
\centering \includegraphics[width=120mm]{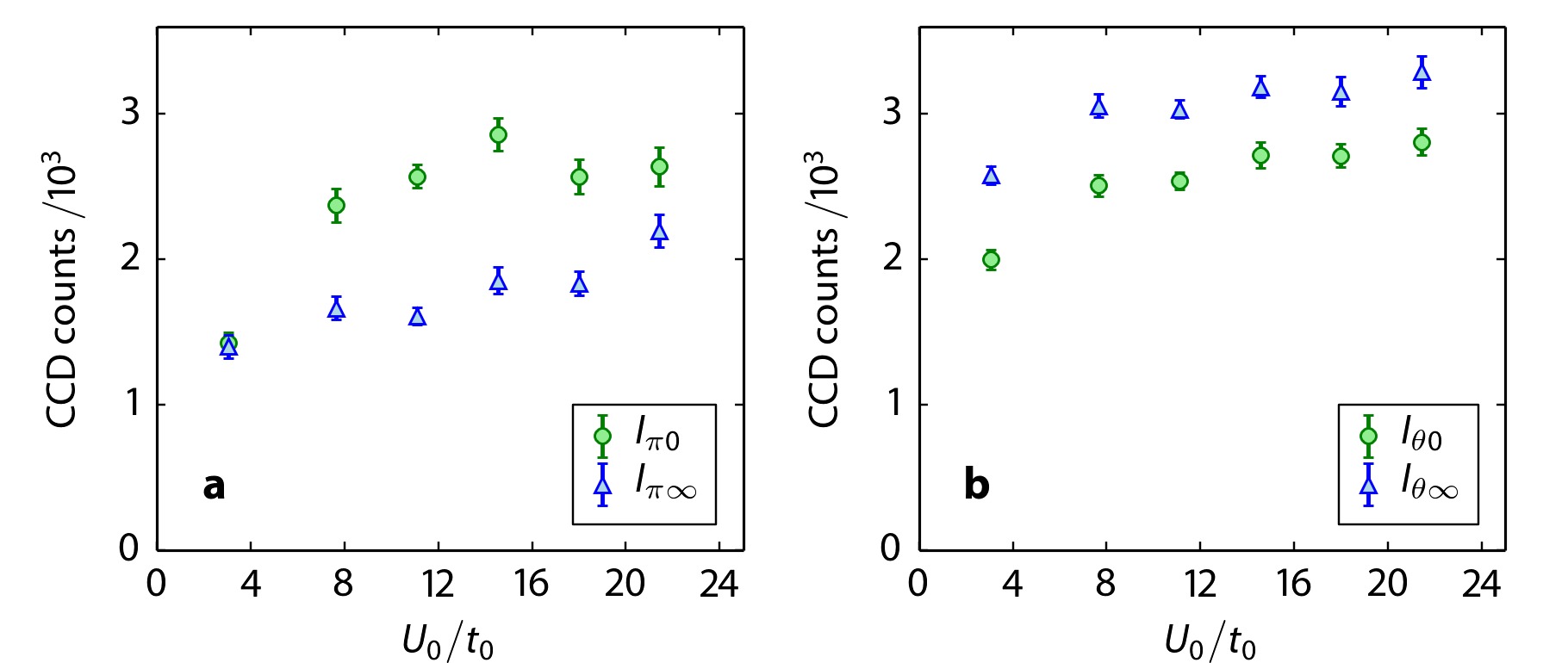}
\caption{\textbf{Detected counts for measurement of
spin structure factor in  Fig.~4.} \textbf{a}, Detected counts vs. 
$U_{0}/t_{0}$, measured for momentum transfer $\mathbfsf{Q}=\bv{\pi}$ for an 
\textit{in-situ} sample ($I_{\bv{\pi}0}$, green circles),  and after decay of 
the Debye-Waller factor, ($I_{\bv{\pi}\infty}$, blue triangles).  As 
$U_{0}/t_{0}$ increases we use a larger atom number to optimise the Bragg 
signal.  $I_{\bv{\pi}\infty}$ and $I_{\bv{\pi}0}$ both increase with 
$U_{0}/t_{0}$ due to the larger $N$,  but $I_{\bv{\pi}0}$ shows an additional 
enhancement due to the presence of AFM correlations.  \textbf{b}, Detected 
counts vs.  $U_{0}/t_{0}$, measured for momentum transfer 
$\mathbfsf{Q}=\bv{\theta}$ for an \textit{in-situ} sample ($I_{\bv{\theta}0}$, 
green circles),  and after decay of the Debye-Waller factor 
($I_{\bv{\theta}\infty}$, blue triangles). For $\mathbfsf{Q}=\bv{\theta}$ most
of the dependence for both the \textit{in-situ} and time-of-flight intensities
is due to the changing $N$.  Error bars in both \textbf{a} and \textbf{b} are
the s.e.m. of at least 40 measurements. The overall count rate is higher for
$\mathbfsf{Q}=\bv{\theta}$ due to the different collection efficiency and gain
settings of the CCD camera. }  
\end{figure}
}
\newcommand{\ExtFigSix}{
\begin{figure}
\centering \includegraphics[width=100mm]{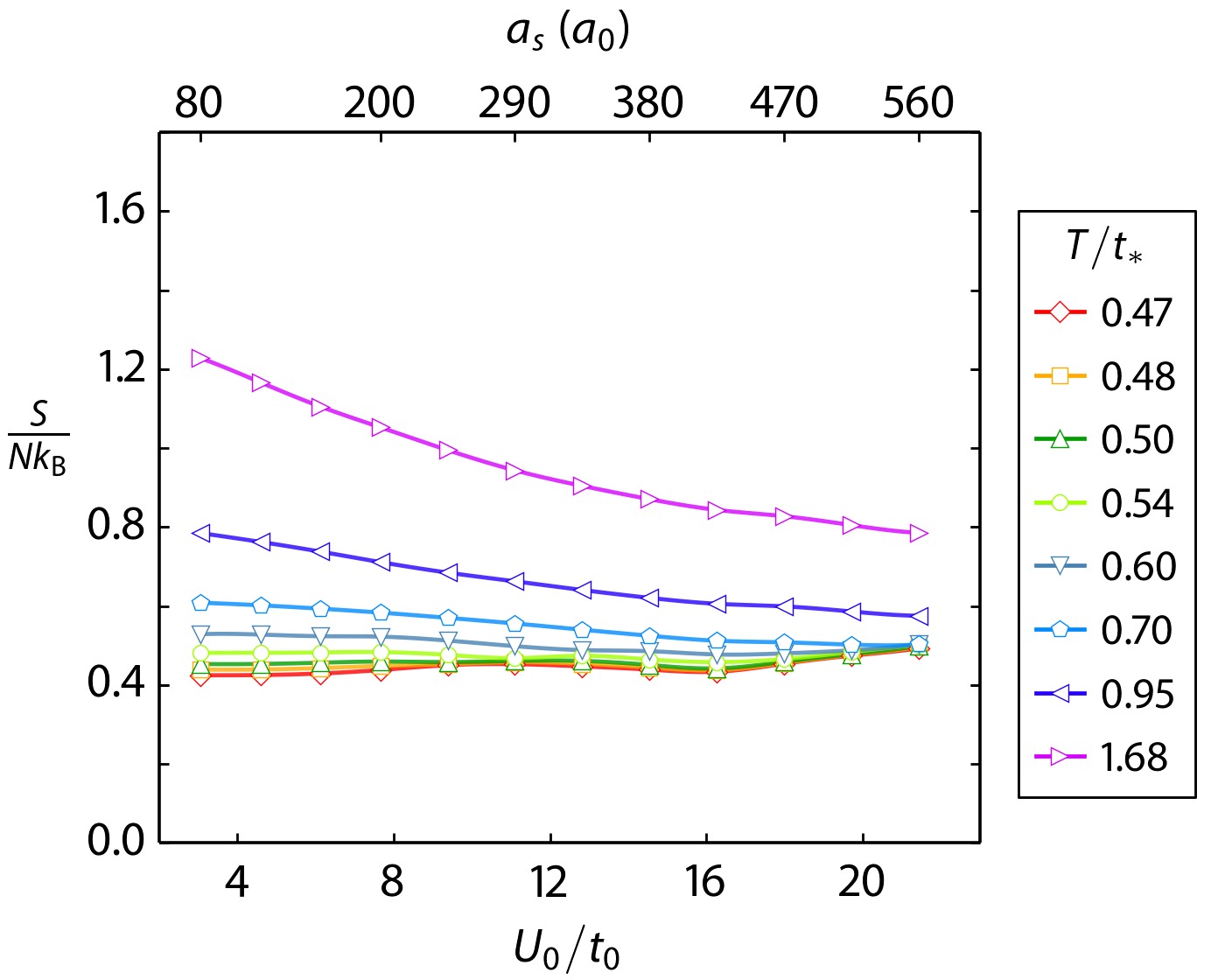}
\caption{\textbf{Entropy per particle at constant
$\bm{T}$.} Overall entropy per particle $S/(Nk_{\mathrm{B}})$ as a function of
$U_{0}/t_{0}$ for the calculations at various $T/t_{*}$ shown in Fig.~4 (lines
are guides to the eye).  For the lowest temperatures, $S/(Nk_{\mathrm{B}})$
does not vary significantly over the range of $U_{0}/t_{0}$ covered by the
experiment justifying the treatment at constant $T$.   A value of
$S/(Nk_{\mathrm{B}})\simeq0.43$ is obtained for the temperature determined from
the data in Fig.~4. This entropy is consistent with $T/T_{F}\simeq 0.04$,
measured in the harmonic dimple trap before loading the atoms into the
lattice.}
\end{figure}
}

\newcommand{\Abstract}{
Ultracold atoms in optical lattices have great potential to contribute to a
better understanding of some of the most important issues in many-body physics,
such as high-temperature (high-$T_c$)
superconductivity~\cite{PhysRevLett.89.220407}.  Thirty years ago, Anderson
suggested that the Hubbard model, a simplified representation of fermions
moving on a periodic lattice, may contain the essence of copper oxide
superconductivity~\cite{ANDERSON87}. The Hubbard model describes many of the
features shared by the copper oxides, including an interaction-driven Mott
insulating state and an antiferromagnetic (AFM) state.  Optical lattices filled
with a two-spin-component Fermi gas of ultracold atoms can faithfully realise
the Hubbard model with readily tunable parameters, and thus provide a platform
for the systematic exploration of its phase diagram~\cite{Jaksch2005,Bloch2008}.
Realisation of strongly correlated phases, however, has been hindered by the
need to cool the atoms to temperatures as low as the magnetic exchange energy,
and also by the lack of reliable thermometry~\cite{McKay2011}. Here we
demonstrate spin-sensitive Bragg scattering of light to measure AFM spin
correlations in a realisation of the three-dimensional (3D) Hubbard model at
temperatures down to 1.4 times that of the AFM phase transition.  This
temperature regime is beyond the range of validity of a simple high-temperature 
series expansion, which brings our experiment close to the limit of the 
capabilities of current numerical techniques. We reach these low temperatures 
using a unique compensated optical lattice technique~\cite{Mathy2012}, in which 
the confinement of each lattice beam is compensated by a blue-detuned laser 
beam. The temperature of the atoms in the lattice is deduced by comparing the 
light scattering to determinantal quantum Monte Carlo~\cite{Blankenbecler1981} 
(DQMC) and numerical linked-cluster expansion~\cite{Rigol2006} (NLCE) 
calculations. Further refinement of the compensated lattice may produce even 
lower temperatures which, along with light scattering thermometry, would open 
avenues for achieving and characterising other novel quantum states of matter, 
such as the pseudogap regime of the 2D Hubbard model.
}

\begin{document}

\title{Observation of antiferromagnetic correlations in the Hubbard model with
ultracold atoms}

\author{Russell A. Hart}
\thanks{These authors contributed equally to this work.} 
\affiliation{Department of Physics and Astronomy and Rice Quantum Institute, Rice
University, Houston, TX 77005, USA}
\author{Pedro M. Duarte}
\thanks{These authors contributed equally to this work.} 
\affiliation{Department of Physics and Astronomy and Rice Quantum Institute, Rice
University, Houston, TX 77005, USA}

\author{Tsung-Lin Yang}
\author{Xinxing Liu} 
\affiliation{Department of Physics and Astronomy and Rice Quantum Institute, Rice
University, Houston, TX 77005, USA} 

\author{Thereza Paiva}
\affiliation{Instituto de Fisica, Universidade Federal do Rio de Janeiro Cx.P.
68.528, 21941-972 Rio de Janeiro RJ, Brazil}

\author{Ehsan Khatami}
\affiliation{Department of Physics, University of California, Davis, California
95616, USA}
\affiliation{Department of Physics and Astronomy, San Jose State University, San
Jose, CA 95192, USA}

\author{Richard T. Scalettar} 
\affiliation{Department of Physics, University of California, Davis, California
95616, USA}

\author{Nandini Trivedi}
\affiliation{Department of Physics, The Ohio State University, Columbus, Ohio 43210,
USA}

\author{David A. Huse}
\affiliation{Department of Physics, Princeton University, Princeton, New Jersey
08544, USA}

\author{Randall G. Hulet}
\email{randy@rice.edu} 
\affiliation{Department of Physics and Astronomy and Rice Quantum Institute, Rice
University, Houston, TX 77005, USA}

\date{\today}

\begin{abstract}
\Abstract
\end{abstract}

\maketitle

A two-spin-component Fermi gas in a simple cubic optical lattice may be
described by a single-band Hubbard model with nearest-neighbour tunnelling $t$
and on-site interaction $U>0$.  At a density $n$ of one  atom per site, and for
sufficiently large $U/t$ there is a crossover from a `metallic' state to a Mott
insulating regime~\cite{RevModPhys.70.1039} as the temperature $T$ is reduced
below $U$.  The Mott regime has been demonstrated with ultracold atoms in an
optical lattice by observing the reduction of doubly occupied
sites~\cite{Jordens2008} and the related reduction of the global
compressibility~\cite{Schneider2008}.  For $T$ below the N\'{e}el ordering
temperature $T_{N}$, which for $U\gg t$  is approximately equal to the exchange
energy $J=4t^{2}/U$, the system undergoes a phase transition to an AFM
state~\cite{Staudt2000}.  In the context of quantum simulations, AFM phases of
Ising spins have been previously engineered with bosonic atoms in an optical
lattice~\cite{Simon2011} and with spin-$\frac{1}{2}$
ions~\cite{Kim2010,Britton2012}.  Also, nearest-neighbour AFM correlations due
to magnetic exchange have been observed along one dimension of an anisotropic
lattice~\cite{Greif2013}.  The same experiment achieved temperatures as low as
$T=0.95t\simeq2.6\,T_N$ when the lattice was configured to be
isotropic~\cite{Imriska2014}, where $T_{N}=0.36t$ is the maximal value of the
N\'{e}el transition temperature~\cite{Staudt2000,Paiva2011,Kozik2013}.

Our experiments are performed with an all-optically produced~\cite{Duarte2011},
quantum degenerate, two-state mixture of the two lowest hyperfine ground states
of fermionic $^{6}$Li atoms,  which we label $|\spup\rangle$ and
$|\spdn\rangle$.  The repulsive interaction between atoms in states
$|\spup\rangle$ and $|\spdn\rangle$ is controlled via a magnetic Feshbach
resonance~\cite{Houbiers1998}, which we use to set the $s$-wave scattering
length $a_{s}$ in the range from 80\,$a_{0}$ to 560\,$a_{0}$, where $a_{0}$ is
the Bohr radius.  A simple cubic optical lattice is formed at the intersection
of three mutually perpendicular infrared (IR) retroreflected laser beams.  We
can dynamically rotate the polarisation of the retroreflection, and thus
continually adjust the potential between a lattice and a harmonic dimple trap.
The overall confinement produced by the Gaussian envelope of each IR lattice
beam is partially compensated with a superimposed, non-retroreflected,
blue-detuned laser beam~\cite{Ma2008,Mathy2012}.  The compensation beams serve
three purposes: (1) They help flatten the confining potential in order to
enlarge the volume of the AFM phase; (2) they provide a way to maintain the
central density near $n\approx1$ as the lattice is loaded; and (3) they may
mitigate the effects of heating in the lattice by lowering the threshold for
evaporation.

A degenerate sample with total atom number $N$ between $1.0 \times 10^{5}$ and
$2.5\times10^{5}$ is prepared in the harmonic dimple trap (without
compensation) at a temperature $T/T_{F}=0.04\pm0.02$, where $T_{F}$ is the
Fermi temperature.  The lattice is turned on slowly to a central depth of
$v_{0}=7\,E_{r}$ (see Methods), where $E_{r}=h^2/(2m\lambda^2)$ is the recoil
energy, $h$ is Planck's constant, $m$ is the atomic mass, and $\lambda=1,064\,
\text{nm}$ is the wavelength of the lattice beams.  While loading the lattice,
the intensities of the compensation beams are adjusted to maintain  a peak
density $n\simeq1$.  We have measured the temperature in the dimple trap before
and after transferring the atoms to the lattice (see Methods and Extended Data
Fig.~3), and have observed that the compensating beams mitigate heating in the
lattice, perhaps by allowing continued evaporative cooling~\cite{Mathy2012} or
by a reduction of three-body loss.

\FigureOne

Bragg scattering of near-resonant
light~\cite{Birkl1995,Weidemueller1998,Miyake2011} is depicted in Fig.~1.  The
Bragg condition for scattering from an AFM ordered sample is satisfied when the
momentum $\bv{Q}$ transferred to a scattered photon is equal to $\bv{\pi}$,
where $\bv{\pi}=\frac{2\pi}{a}\hhh$ is a reciprocal lattice
vector of the magnetic sublattice, and $a=\lambda/2$ is the lattice spacing.
Cameras are positioned to detect scattering at $\bv{Q}=\bv{\pi}$ and also at
$\bv{Q}=\bv{\theta}$, a momentum transfer that does not satisfy the Bragg
condition and is used as a control.  We obtain spin sensitivity, in analogy to
neutron scattering in condensed matter, by setting the Bragg laser frequency
between the optical transition frequencies for the two spin
states~\cite{Corcovilos2010,Werner2005}.  Prior to the measurement, we jump
$v_{0}$ to $20\,E_{r}$ in a few $\mu\text{s}$ to lock the atoms in place (see
Methods), and then illuminate them \textit{in-situ} for $1.7\mu\text{s}$ with
the Bragg probe.  Alternatively, we can suddenly turn off the $20\,E_{r}$
lattice and illuminate the atoms after time-of-flight $\tau$.

\FigureTwo

Figure~2 shows the results of simultaneous measurements of the scattered
intensity for $\bv{Q}=\bv{\pi}$ and $\bv{Q}=\bv{\theta}$ ($I_{\bv{\pi}}$ and
$I_{\bv{\theta}}$, respectively), as a function of $\tau$.  After a few
$\mu\text{s}$ of expansion, when the extent of the atomic wavepackets becomes
comparable to the lattice spacing, the light scattered from correlated spins no
longer interferes constructively at the detector.  More precisely, the
Debye-Waller factor $e^{-2W_{\bv{Q}}(\tau)}=
\exp\left[-\sum_{i=x,y,z}Q_{i}^{2}\expect{r_{i}^{2}}_{\tau}\right]$ decays to
zero after a sufficiently long $\tau$ (see Methods) and the sample is
effectively uncorrelated.  Here $r_{i}$ is the displacement of an atom from the
centre of the lattice site at which it was initially localised.

By comparing the intensity of the light scattered \textit{in-situ} ($\tau=0$)
to that after sufficiently long $\tau$ ($I_{\bv{Q}0}$ and $I_{\bv{Q}\infty}$,
respectively), we effectively normalise the Bragg scattering signal to the
diffuse scattering background of an uncorrelated sample, achieving high
sensitivity to magnetic ordering and strong rejection of common mode
systematics.  Figure~2 shows that there is enhanced scattering at $\tau=0$
relative to the uncorrelated cloud ($\tau=9\,\mu\text{s}$) for
$\bv{Q}=\bv{\pi}$, whereas for $\bv{Q}=\bv{\theta}$ scattering at $\tau=0$ is
reduced, such that $I_{\bv{\theta}0}/I_{\bv{\theta}\infty}<1$.  Double
occupancies, present as `virtual' states even at low
temperatures~\cite{Fuchs2011}, reduce coherent scattering in all directions,
since each atom in the pair has opposite spin and therefore scatters with
opposite phase.  For $\bv{Q}=\bv{\pi}$ the coherent enhancement from  AFM spin
correlations exceeds this reduction. Furthermore, the coherent enhancement of
the signal along $\bv{Q}=\bv{\pi}$  suppresses the scattered intensity in other
directions.

For a momentum transfer $\bv{Q}$, the spin structure factor $S_{\bv{Q}}$
of the sample is defined as
\begin{equation}
S_{\bv{Q}} \equiv \frac{4}{N} \sum_{i,j} e^{i\bv{Q}\cdot ( \bv{R}_{i} -
\bv{R}_{j} )} \expect{\sigma_{zi}\sigma_{zj}}  
\end{equation}
Here $N$ is the total number of atoms, the sums extend over all lattice sites
$i$ and $j$, $\bv{R}_{j}$ is the location of the $j^{\text{th}}$ site, and
$\sigma_{zj}$ is the $z$ component of the spin operator for the $j^{\text{th}}$
site: 
\begin{eqnarray*}
 \sigma_{zj}
\ket{0}_{j}=0\ket{0}_{j} &  ~~~~~~~~~~ & \sigma_{zj} |\spup\rangle_{j} =
+ \frac{1}{2} |\spup\rangle_{j}  \\
\sigma_{zj} |\spdn\rangle_{j} =  -
\frac{1}{2} |\spdn\rangle_{j} & ~~~~~~~~~~ & \sigma_{zj} |\dbl\rangle_{j} =
0|\dbl\rangle_{j} 
\end{eqnarray*}
In a sample with complete AFM ordering $S_{\bv{\pi}}\simeq\,\!N$, whereas for
uncorrelated samples in the lattice $S_{\bv{\pi}}\leq1$ and
$S_{\bv{\theta}}\leq1$.  The choice of the $z$ spin component for this analysis
is arbitrary, as each of the other axes would result in the same value for
$S_{\bv{Q}}$ in the absence of a symmetry-breaking field.  In the limit of
tightly localised wavefunctions ($e^{-2W_{\bv{Q}}(\tau=0)}\approx1$), and for a
weak probe, the spin structure factor is $S_{\bv{Q}}\approx
I_{\bv{Q}0}/I_{\bv{Q}\infty}$.  We determine the spin structure factor by
measuring the scattered intensities $I_{\bv{Q}0}$ and $I_{\bv{Q}\infty}$ and
applying a correction to account for the \textit{in-situ} Debye-Waller factor
in the 20\,$E_{r}$ lattice and for saturation of the atomic transition, which
generates a small component of inelastically scattered light (see Methods).  

Within the local density approximation (LDA) we model the sample by considering
each point in the trap as a homogeneous system in equilibrium at a temperature
$T$,  with local values of the chemical potential and the Hubbard parameters
determined by the trap potential.  The spin structure factor of the sample
$S_{\bv{Q}}$ can then be expressed as the integral over the trap of the local
spin structure factor per lattice site, $s_{\bv{Q}}$.  Figure~3a shows
numerical calculations of $s_{\bv{\pi}}$ for various temperatures in a
homogeneous lattice with $U/t=8$, close to where $T_{N}$ is
maximal~\cite{Staudt2000}.  The figure shows that $s_{\bv{\pi}}$ is sharply
peaked around $n=1$ and grows rapidly as $T$ approaches $T_{N}$ from above. 

\FigureThree

Figures~3b and 3c show $n$ and $s_{\bv{\pi}}$ profiles, respectively,
calculated for our experimental parameters at various values of $U_{0}/t_{0}$,
where $U_{0}$ and $t_{0}$ denote the local values of the Hubbard parameters at
the centre of the trap.   As seen in Fig.~3b,  only a fraction of the atoms in
the sample is near $n=1$, where AFM correlations are maximal.  The finite
extent of the lattice beams causes the lattice depth to decrease with distance
from the centre, resulting in an increasing $t$ such that both $U/t$ and $T/t$
decrease with increasing radius for constant $T$ (see Extended Data Fig.~1).
The radial decrease in $T/t$ causes $s_{\bv{\pi}}(r)$  to maximise at the
largest radius for which the density is $n\approx1$.   For large $U_{0}/t_{0}$
the cloud exhibits an $n=1$ Mott plateau and $s_{\bv{\pi}}(r)$ is maximised at 
the outermost radius of the plateau.

In the experiment, we measure $S_{\bv{Q}}$ as a function of $U_{0}/t_{0}$.  At
each value of $U_{0}/t_{0}$ we vary the atom number $N$ to maximise the
measured $S_{\bv{\pi}}$ (see Methods and Extended Data Fig. 2).  According to
the picture presented above,  this has the effect of optimising the size and
location of the $n=1$ region of the cloud such that AFM correlations are
maximised.   The compensation strength $g_{0}$, which is the same for all
$U_{0}/t_{0}$, was also adjusted to maximise $S_{\bv{\pi}}$.  We found the
optimum to be $g_{0}=3.7\,E_{r}$ at a lattice depth $v_{0}=7\,E_{r}$ (see
Methods).   Besides the equilibrium considerations regarding the optimal size
and location of the Mott plateau, we believe that the dynamical adjustment of
$g_{0}$ during the lattice turn-on reduces the time for the system to
equilibrate, by minimising the deviation of the equilibrium density
distribution in the final potential from the starting density distribution in
the dimple trap prior to loading the lattice.

Figure~4 shows the measured values of $S_{\bv{\pi}}$ and $S_{\bv{\theta}}$ at
optimal $N$ for various values of $U_{0}/t_{0}$ (see Extended Data Fig.~5
for the raw counts at the CCD cameras).  We find that $S_{\bv{\pi}}$ is peaked
for $11 < U_{0}/t_{0} < 15$.  In contrast, the measurements of
$S_{\bv{\theta}}$ vary little over the range of interaction strengths,
consistent with an absence of coherent Bragg scattering in this direction.
Measurements of $S_{\bv{\pi}}$ after hold time in the lattice show that the
Bragg signal decays for larger temperatures (see Extended Data Fig.~4).
Comparing the measured $S_{\bv{\pi}}$ with numerical calculations for a
homogeneous lattice (for example, those in Fig.~3a) allows us to set a trap
independent upper limit on the temperature, which we determine to be
$T/t_{0}<0.7$.  

Precise thermometry is obtained by comparing the measured $S_{\bv{\pi}}$ with
numerical calculations averaged over the trap density distribution for
different values of $T$.  The results of such numerical calculations are shown
in Fig.~4, labelled by the value of $T/t_{*}$, which we define as the local
value of $T/t$ at the radius where the spin structure factor per lattice site
is maximal (see Fig.~3c).  At $U_{0}/t_{0}=11.1$, where measured AFM
correlations are maximal, we find $T/t_{*}=0.51\pm0.06$,  where the uncertainty
is due to the statistical error in the measured $S_{\bv{\pi}}$ and the
systematic uncertainty in the lattice parameters used for the numerical
calculation.  This temperature is consistent with the data at all values of
$U_{0}/t_{0}$.  We caution, however, that for values of $U/t>10$  a single-band
Hubbard model may not be adequate, as corrections involving higher bands may
become non-negligible~\cite{Werner2005,Mathy2009}.

\FigureFour
 
As was shown in Fig.~3b,  for $U_{0}/t_{0}=11.1$ the dominant contribution to
$S_{\bv{\pi}}$ comes from the outermost radius of the Mott plateau. At that
radius, the local value of $U/t$ is $U_{*}/t_{*}=9.1$, consistent with DQMC
calculations for the homogeneous lattice~\cite{Staudt2000,Paiva2011,Kozik2013},
which find $T_{N}$ to be maximised for $U/t$ between 8 and 9.  For
$U_{0}/t_{0}=11.1$, $t_{*}=1.3\,\mathrm{kHz}$, so we can infer the temperature
of the system to be $T=32\pm4\,\mathrm{nK}$.   In terms of $T_{N}$, the
temperature is $T/T_{N}=1.42\pm0.16$.   At this temperature, the numerical
calculations indicate that the correlation length is approximately the lattice
spacing.  The calculations show that the entropy per particle in the trap is
$S/(Nk_{\mathrm{B}})\simeq0.43$, where $k_{\mathrm{B}}$ is the Boltzmann
constant (see Extended Data Fig.~6).  This entropy range is consistent
with $T/T_{F}=0.04\pm0.02$ measured in the harmonic dimple trap before loading
the atoms into the lattice~\cite{Kohl2006}, and thus justifies the assumption of
adiabatic loading.  

In conclusion, we have observed AFM correlations in the 3D Hubbard model using
ultracold atoms in an optical lattice via spin-sensitive Bragg scattering of
light.  Because magnetic order is extremely sensitive to $T$ in the vicinity of
$T_{N}$, Bragg scattering provides precise thermometry in regimes previously
inaccessible to quantitative temperature measurements.  While previous cold
atom experiments on the 3D Fermi-Hubbard model were in a temperature regime
that could be accurately represented by a simple high-temperature series
expansion, the data presented here is near the limit of the capabilities of
advanced numerical simulations.  Our experimental setup can be configured to
study the 2D Hubbard model in an array of planes; further progress to lower
temperature will put us in a position to answer questions about competing
pairing mechanisms in 2D, and may ultimately resolve the long standing question
of d-wave superconductivity in the Hubbard model.

%\bibliography{afm}

%%==============================================================================

\section{Acknowledgements}
This work was supported under ARO Grant No.\,W911NF-13-1-0018 with funds
from the DARPA OLE program,  NSF, ONR, the Welch Foundation (Grant No. C-1133),
and an ARO-MURI grant No.\,W911NF-14-1-003.  T.P. acknowledges support from
CNPq, FAPERJ, and the INCT on Quantum Information. RTS acknowledges support 
from the Office of the President of the University of California.

\section{Author Contributions} The experimental work was performed by R.A.H.,
P.M.D., T.-L.Y., X.L., and R.G.H., while T.P., E.K., P.M.D., R.T.S., N.T., and
D.A.H. performed the theory needed to extract temperatures from the data and
provided overall theoretical guidance.  All authors contributed to the writing
of the manuscript.
\appendix*
\renewcommand{\theequation}{\arabic{equation}}
\setcounter{figure}{0}
\def\figurename{}
\renewcommand{\thefigure}{EXTENDED DATA FIG. \arabic{figure}}

%%==============================================================================

\section{METHODS}

\subsection{Preparation.} $^6$Li atoms are first captured and cooled in a
magneto-optical trap (MOT) operating at 671 nm. They are further cooled in a
second MOT stage employing 323 nm light near resonant with the
$2S\hspace{-0.0pt}\rightarrow\hspace{-0.0pt}3P$ transition.  As described
previously~\cite{Duarte2011},  these atoms are laser cooled into a large volume
optical dipole trap (ODT) where a balanced spin mixture of the states
$|\spup\rangle=|2S_{1/2};F=1/2,m_{F}=+1/2\rangle$ and
$|\spdn\rangle=|2S_{1/2};F=1/2,m_{F}=-1/2\rangle$ is produced.
 
Once the large volume ODT is loaded, we set the magnetic field to 340~G
($a_{s}\simeq-289\,a_{0}$) to perform evaporative cooling.  The intensities of
the lattice beams (1,064\,nm) in dimple configuration (with the polarisation of
each retroreflection perpendicular to that of each input beam) are turned on in
1\,s.  The depth of the dimple, which at this point is only a small
perturbation on the ODT, is adjusted to control the final atom number in the
experiment.  The depth of the ODT is then ramped to zero in 5.5\,s to
evaporatively cool the atoms into the dimple.  To produce a final sample with
repulsive interactions, the magnetic field is increased to 595\,G
($a_{s}\simeq326\,a_{0}$) in a 5\,ms linear ramp starting 3\,s into the
evaporation trajectory.   Due to the small volume of the dimple relative to the
ODT, evaporation into the dimple is efficient and deeply degenerate samples are
reliably produced.   

We measure $T/T_{F}$ in the dimple trap by fitting the density profile, after
0.5~ms of time-of-flight, to a Thomas-Fermi distribution~\cite{Butts1997}.  The
magnetic field is tuned to 528~G to make the gas non-interacting prior to the
measurement.  For the experiments reported here, the final dimple depths are in
the range between $0.325\,E_r$ and $0.5\,E_r$ per axis, resulting in $N$
between $1.0-2.5\times10^{5}$.  The measured value $T/T_{F}=0.04\pm0.02$ is
independent of $N$ within this range.  The uncertainty in $T/T_{F}$ is the
standard deviation of the fitted value for at least six independent
realisations.

\subsection{Compensated optical lattice.} The experiment takes place in a
compensated simple cubic optical lattice potential that can be expressed as
\begin{equation*}
  V_{3D}(x, y, z)=V_{1D}( x; y,z)+V_{1D}( y; z,x)+V_{1D}(z; x,y),
\end{equation*}
where
\begin{equation*}
  V_{1D}( x; y ,z )=V_{L}(x; y,z)+V_{C}(x; y, z),
\end{equation*}
and $V_{L}$, $V_{C}$ are the potentials produced by the lattice (1,064\,nm) and
compensation (532\,nm) beams respectively:
\begin{equation*}
  V_{L}( x; y,z)  =
  - v_{0} \exp\left[- 2 \frac{ y^{2} + z^{2} }{w_{L}^{2}} \right]
  \cos^{2}( \frac{2\pi}{\lambda} x )
\end{equation*}
\begin{equation*}
  V_{C}( x; y, z)  =
   g_{0}  \exp\left[ -2 \frac{ y^{2} + z^{2} }{w_{C}^{2}} \right].
\label{eq:Vcomp}
\end{equation*}
Here, $v_{0}$ is the lattice depth and $g_{0}$ is the compensation
($v_{0}\text{,}\,g_{0}>0$).   A schematic of the compensated lattice, and the
spatial variation of the Hubbard parameters due to the finite lattice beam
waists, are shown in Extended Data Fig.~1. 

The beam waists ($1/e^{2}$ radius) of the three axes are calibrated
independently by phase modulation spectroscopy of each lattice beam and by
measuring the frequency of breathing mode oscillations. The waists are found to
be (up to a $\pm5\%$ systematic uncertainty) $w_{L}=(47,47,44)\,\mu$m and
$w_{C}=(42,41,40)\,\mu$m,  for beams propagating along $x,y,z$, respectively.

\subsection{Lattice loading.} To load the lattice from the dimple trap we first
rotate the polarisation of the retroreflected beams parallel to that of the
input beams in 100 ms.  In the following 25~ms, we increase the lattice
depth to $2.5\,E_r$ and ramp the magnetic field to set the final value of
$U_{0}/t_{0}$.  The lattice depth is then ramped to $7.0\,E_r$ in 15~ms.

Throughout the process of loading the lattice from the dimple, the power of the
compensating beams is adjusted in order to maintain the peak density of the
sample at $n\simeq1$.  At the final lattice depth of $v_{0}=7.0\,E_r$, the
average compensation per beam is $g_{0}=3.7\,E_{r}$.  The value of $g_{0}$ for
each beam is adjusted slightly from this average in order to create samples
that appear spherically symmetric.   

\subsection{Round-trip $\bm{T/T_{F}}$ measurements.}  After loading the atoms
into the $7E_{r}$ lattice we wait for a hold time $t_{h}$ and then reverse the
lattice loading ramps to return to the harmonic dimple trap and measure
$T/T_{F}$.  This measurement, shown in Extended Data Fig.~3, sets an upper
limit on the entropy of the system in the lattice,  and is also a measure of
the heating rate of the system in the lattice. 

\subsection{Temperature dependence of $S_{\bv{\pi}}$} 

In Extended Data Fig.~4 we show $S_{\bv{\pi}}$ as a function of hold time in
the lattice $t_{h}$ and observe that it decays for longer hold times, as
expected from  the increase in $T/T_{F}$.   Although the preparation of the
sample and the final potential are somewhat different for the data presented in
Extended Data Figs.~3 and 4, the data support the contention that the Bragg
signal decreases with increasing $T$.

\subsection{Variation of $\bm{N}$ to maximise $\bm{S_{\pi}}$.}  The global
chemical potential $\mu_{0}$ must be increased for larger $U_{0}/t_{0}$ to
guarantee the formation of a Mott plateau in the trap.   A larger $\mu_{0}$
results in larger atom number.  $N$ is adjusted to maximise the Bragg signal
for each experimental value of $U_{0}/t_{0}$ in Fig.~4.  We adjust $N$ by
tuning the depth of the dimple trap in which degeneracy is achieved prior to
loading the atoms into the lattice.  The optimal value of $N$ as a function of
$U_{0}/t_{0}$ is shown in Extended Data Fig.~2. 

\subsection{Spin structure factor measurement.} We measure the spin structure
factor at two different values of the momentum transfer $\bv{Q}$ given by
\begin{align*}
\bv{\pi} & =\frac{2\pi}{a}(-0.5,-0.5,+0.5) \\
\bv{\theta} & =\frac{2\pi}{a}(+0.396,-0.105,-0.041),
\end{align*}  
where $a=\lambda/2$ is the lattice spacing. 

We detect the scattered light using two separate cameras as the cloud is
illuminated with the Bragg probe beam for $1.7\,\mu$s.  The Bragg probe beam is
a collimated Gaussian beam with a waist of $450\,\mu$m and 
$250\,\mu$W of power, resulting in an intensity
$I_{\mathrm{p}}=79\,\mathrm{mW}/\mathrm{cm}^{2}$.   The intensity of the probe
determines the on-resonance saturation parameter  $s_{0}=I_{\text{p}}|
\bv{\hat{e}}_{\text{p}}\cdot\bv{\hat{e}}_{-1}|^{2}/\left(\frac{\pi
hc\Gamma}{3\lambda_{0}^{3}}\right)=15.5$, where $c$ is the speed of
light, $\bv{\hat{e}}_{\text{p}}$ is the polarisation of the probe light,
$\bv{\hat{e}}_{-1}$ is the unit vector in the direction of the dipole matrix
element of the transition, $\lambda_{0}=671\,\mathrm{nm}$ is the wavelength of
the transition, and $\Gamma$ is its linewidth.  The polarisation of the
incident light in our experiment is linear and perpendicular to the
quantisation axis, so $| \bv{\hat{e}}_{\text{p}}\cdot
\bv{\hat{e}}_{-1}|^{2}=1/2$.  The Bragg probe detuning is set between the two
spin states, such that $\Delta=|\Delta_{\ \spup}|=|\Delta_{\
\spdn}|=6.4\Gamma$, where $\Delta_{\ \spup}$ and $\Delta_{\ \spdn}$ are the
detunings from the two spin states. 

The spin structure factor is defined in equation (1) as a sum over lattice
sites $i,\,j$.  By quickly ramping the lattice depth to $v_{0}=20\,E_{r}$, the
state of the system is projected into a product state, where the wavefunction
of each atom is localised at a lattice site. Hence, we can write $S_{\bv{Q}}$
as a sum over particles $m,n$:
\begin{equation*}
    S_{\bv{Q}} = \frac{4}{N} \sum_{m,n}
     e^{i\bv{Q}\cdot ( \bv{R}_{m} - \bv{R}_{n} )}
      \langle \sigma_{z} \rangle_{m} \langle \sigma_{z}  \rangle_{n}
\label{eq:sqdef}
\end{equation*}
where $\langle\sigma_{z}\rangle_{n}$ is the $z$ component of the spin of the
$n^{\text{th}}$ atom.

When illuminated with the probe light, each atom can be considered as an
independent scatterer, and the intensity at the detector can be obtained by
summing the field contributions from the individual atoms and squaring the
total field.   We assume that the spatial wavefunction of all atoms is the
harmonic oscillator ground state in a lattice site of depth $v_{0}$, and that
it does not change during the measurement.  The resulting intensity at the
detector is given by 
\begin{multline}
  I_{\bv{Q}}(\tau) = 
   \frac{ A s_{0}/2}{ 4\delta^{2} + s_{0} } N  \\
   + 
  e^{-2W_{\bv{Q}}(\tau)}
  \frac{ 2 A s_{0} \delta^{2} }{ (4\delta^{2} + s_{0} )^{2} }  
   \sum_{\genfrac{}{}{-2pt}{}{m,n}{m\neq n}} 
  4 \langle \sigma_{z} \rangle_{m}  \langle \sigma_{z} \rangle_{n}
  e^{ i \bv{Q} \cdot ( \bv{R}_{n} - \bv{R}_{m} ) }   
\end{multline}
where $\delta=\Delta/\Gamma$, and
$A=\frac{3}{8\pi}\frac{\hbar ck\Gamma}{r_{D}^{2}}|\bv{\Lambda}|^{2}$.  Here
$\bv{\Lambda}$ is the polarisation vector of the scattered field, $\bv{\Lambda}
= \hat{\bv{n}} \times ( \hat{\bv{n}} \times \bv{\hat{e}}_{-1})$, where
$\hat{\bv{n}}$ is a unit vector pointing in the direction of the detector.

In equation (2) the  first term arises from uncorrelated scattering by the
atoms, while the second term represents the interference due to magnetic
correlations. We can identify the spin structure factor in the interference
term as 
\begin{equation*} 
\sum_{\genfrac{}{}{-2pt}{}{m,n}{m\neq n}}
4 \langle \sigma_{z} \rangle_{m}  \langle \sigma_{z} \rangle_{n} e^{ i \bv{Q}
\cdot ( \bv{R}_{n} - \bv{R}_{m} ) }   = N(S_{\bv{Q}} -1 ) 
\end{equation*} 
and obtain 
\begin{equation*}
S_{\bv{Q}} = 1 + C_{\bv{Q}}(\tau) \left(\frac{ I_{\bv{Q}}(\tau) }{
I_{\bv{Q}\infty} } -1 \right) 
\end{equation*}
where $ I_{\bv{Q}\infty} =  \frac{ A s_{0}/2}{ 4\delta^{2} + s_{0} } N $, and
the correction factor is
$C_{\bv{Q}}(\tau)=e^{2W_{\bv{Q}}(\tau)}(1+\frac{s_{0}}{4\delta^{2}})$.  In the
experiment we obtain $S_{\bv{Q}}$ by combining measurements of the scattered
intensity \textit{in-situ} ($\tau=0$) and after sufficiently long
time-of-flight ($\tau=6\,\mu\mathrm{s}$).   The correction factor takes the
values $C_{\bv{\pi}}(\tau=0)=1.52$ for $\bv{Q}=\bv{\pi}$ and
$C_{\bv{\theta}}(\tau=0)=1.18$ for $\bv{Q}=\bv{\theta}$.

\subsection{Time-of-flight.} After the atoms are released in time-of-flight, the
Debye-Waller factor decays as the atomic wavefunctions expand, resulting in a
corresponding decay of the Bragg scattered intensity.   For a lattice of depth
$v_{0}$ 
\begin{equation*}
  e^{-2W_{\bv{Q}}(\tau)} =  e^{-2W_{\bv{Q}}(\tau=0)} \exp \left[ -
\frac{\sqrt{v_{0}/E_{r}}}{2} \left(  \frac{ |\bv{Q}| h }{ 2ma } \right)^{2}
\tau^{2} \right]. 
\end{equation*}
This equation was used to calculate the solid grey line in Fig.~2.  The average
value of the Debye-Waller factor during the duration of the Bragg exposure
\begin{equation*}
(1.7\,\mu\mathrm{s})^{-1}\int_{\tau}^{\tau+1.7\,\mu\mathrm{s}}
e^{-2W_{\bv{Q}}(\tau')} \mathrm{d}\tau'
\end{equation*}
is used to calculate the dashed grey line in Fig.~2.  

The data shown in Fig.~2 was taken at $U_{0}/t_{0}=13.4$ with
$N=2.5\times10^{5}$ atoms. This value of $N$ is above the optimal value, so the
ratio of $I_{\bv{\pi}0}/I_{\bv{\pi}\infty}$ in Fig.~2 gives $S_{\bv{\pi}}\simeq
1.4$, which is less than the expected optimal value of $S_{\bv{\pi}}$ from
Fig.~4. 

\subsection{Momentum transferred from the probe to the atoms.} As  mentioned
above,  we assume that the spatial wavefunction of the atoms remains unchanged
for the duration of the exposure.  For this assumption to be valid, the
Lamb-Dicke  parameter
$\eta^{2}=\frac{h^{2}/(2m\lambda_{0}^{2})}{2E_{r}\sqrt{v_{0}/E_{r}}}$
needs to be $\ll1$.
In the 20\,$E_{r}$ lattice, $\eta^{2}=0.27$, meaning that approximately one
out of every 4 photons scattered will excite an atom to the second band of the
lattice.    An atom in the second band has larger position uncertainty and
hence a smaller Debye-Waller factor,  which reduces its contribution to the
Bragg scattering signal.

The total number of photons scattered per atom is given by
$N_{\text{p}}=t_{\text{exp}}\Gamma\frac{s_{0}/2}{s_{0}+4\delta^{2}}$,  where
the duration of the probe is $t_{\text{exp}}=1.7\,\mu$s. For $s_{0}=15.5$ and
$\delta=6.4$, $N_{\text{p}}=2.7$, thus justifying the assumption that
the atoms remain in the lowest band during the pulse. 

For the Bragg scattering measurements performed after time-of-flight, the
momentum transferred from the probe to the atoms plays a more significant role,
since the atoms are not trapped and will recoil after every photon scatter.
Despite this, we still see a good agreement between the observed decay of the
Bragg scattering signal and the decay expected for a Heisenberg limited
wavepacket, as shown in Fig.~1.   We have also performed non-spin-sensitive
Bragg scattering measurements from the 010 planes of the lattice and observe
the same agreement,  justifying that momentum transfer from the probe to the
atoms can be neglected for the exposure times used.

\subsection{Optical density.} A low optical density of the sample is important
so that the probe is unattenuated through the atom cloud, and multiple
scattering events of the Bragg scattered photons are
limited~\cite{Corcovilos2010}.  The optical density  can be approximated as
\begin{equation*}
   \text{OD}
  \simeq
    \frac{ \sigma_{0} | \bv{\hat{e}}_{\text{p}} \cdot \bv{\hat{e}}_{-1} |^{2}}
         {  4 \delta^{2} + s_{0} }
    \frac{1}{a^{2}}
    \left(\frac{ 3N}{4\pi} \right)^{1/3}
\end{equation*}
where $\sigma_{0}=3\lambda_{0}^{2}/2\pi$.  With $s_{0}=15.5$,
$\delta=6.4$  and $N=1.8\times10^{5}$ atoms we have
$\text{OD}\simeq0.072$\,.  At this value we do not expect significant
corrections to the spin structure factor measurement due to the attenuation of
the probe.  We have not included any corrections in our measurement due to
finite optical density effects.  

\subsection{Light collection.} We collect Bragg scattered light in the
$\bv{\pi}$ direction over a full angular width of  110~mrad,  given by a 2.5 cm
diameter collection lens located 23 cm away from the atoms.    In the
$\bv{\theta}$ direction, light is collected by a 2.5 cm diameter  lens placed 8
cm away from the atoms, corresponding to a full angular width of 318~mrad.  The
scattered light in each of the directions is focused to a few pixels on the
cameras, so no additional angular information is obtained.  For
$N=1.8\times10^{5}$, $s_{0}=15.5$, $\Delta=6.4\,\Gamma$ and a
$1.7\,\mu\text{s}$ pulse, the detector in the $\bv{\pi}$ direction collects
approximately 1300 photons, whereas the detector in the $\bv{\theta}$ direction
collects approximately $10^4$ photons.   The noise floor from readout, dark
current and background light per shot has a variance equivalent to
approximately 250 photons in the $\bv{\pi}$ direction and 1000 photons in the
$\bv{\theta}$ direction.

\subsection{Data averaging.} The signals we detect are small enough that an
uncorrelated sample may, in a single shot, produce a scattering signal as large
as the ones produced by samples with AFM correlations.  To obtain a reliable
measurement of $S_{\bv{\pi}}$ we average at least 40 \textit{in-situ} shots to
obtain $I_{\bv{Q}0}$ and at least 40 time-of-flight shots to obtain
$I_{\bv{Q}\infty}$.

We estimate the expected variance on $S_{\bv{\pi}}$ by considering a randomly
ordered sample in which
$e^{i\bv{\pi}\cdot\bv{R}_{n}}2\langle\sigma_{z}\rangle_{n} $  is equal to +1 or
-1 with equal probability.  $S_{\bv{\pi}}$ can be written as
\begin{equation*}
  S_{\bv{\pi}} =
    \left| \sum_{n} e^{i\bv{\pi}\cdot \bv{R}_{n}}
    \frac{2 \langle \sigma_{z}  \rangle_{n} }{\sqrt{N}}  \right|^{2}, 
\end{equation*}
which is equivalent to the square of the distance travelled on an unbiased
random walk with step size $1/\sqrt{N}$.   The mean and standard deviation can
then be readily calculated: $\overline{S_{\bv{\pi}}}=1$ and
$\sqrt{\text{Var}(S_{\bv{\pi}})}=\sqrt{2}$, where $\text{Var}(S_{\bv{\pi}})$
denotes the variance of the random variable $S_{\bv{\pi}}$.  With a standard
deviation that is larger than the mean value, a considerable number of shots
needs to be taken in order to obtain an acceptable error in the mean.    The
standard error of the mean for 40 shots will be $\sqrt{2/40}=0.22$, consistent
with what we obtain in the experiment (see Fig.~4).

\subsection{Numerical calculations.}  DQMC and NLCE calculations are used to
obtain the local values of the thermodynamic quantities in our trap, including
the density, entropy, and the spin structure factor.   DQMC calculations for
arbitrary chemical potential (and hence density)  can be obtained reliably down
to temperatures slightly above the N\'{e}el temperature for a given
$U/t\lesssim9$.   For stronger interactions intermediate values of $n$ become
inaccessible to DQMC due to the sign problem, in which case we rely on the NLCE
to obtain values of the thermodynamic quantities for arbitrary chemical
potential down to temperatures as low as $T/t=0.40$.  

DQMC results for a \parbox[b]{2.7em}{$6\!\!\times\!\!6\!\!\times\!\!6$} lattice
were obtained with the methodology described in Refs.~\cite{Blankenbecler1981}
and \cite{Paiva2010}.  Inverse temperature discretisation $\Delta \tau =
\beta/L=1/20t$ is sufficiently small that Trotter corrections are substantially
less than statistical error bars.  Finite size effects were assessed by
comparing DQMC results for \parbox[b]{2.7em}{$6\!\!\times\!\!6\!\!\times\!\!6$}
and \parbox[b]{2.7em}{$8\!\!\times\!\!8\!\!\times\!\!8$} lattices.  Differences
are only appreciable when the spin structure factor per lattice site,
$s_{\bv{\pi}}>5$.  The local value of $s_{\bv{\pi}}$ is always less than 4 in
our calculations, so DQMC results in a
\parbox[b]{2.7em}{$6\!\!\times\!\!6\!\!\times\!\!6$} lattice are sufficient for
the comparison with theory.  

In NLCEs, an extensive property of the lattice model per site in the
thermodynamic limit is expressed in terms of contributions from finite clusters
that can be embedded in the lattice. NLCEs use the same basis as
high-temperature expansions, however, properties of clusters are calculated via
exact diagonalisation, as opposed to a perturbative expansion in powers of the
inverse temperature~\cite{Rigol2006,Tang2013}. The site-based NLCE for the
Hubbard model~\cite{Khatami2011} is implemented here for a three-dimensional
lattice and carried out to the eighth order for all thermodynamic quantities,
except for $S_{\bv{\theta}}$, where due to the reduced symmetry, only seven
orders were obtained.  Within its region of convergence ($T/t\gtrsim 1.5$ for
any $n$ and $U$), NLCE results do not contain any systematic or statistical
errors. The convergence region extends to significantly lower $T/t$ at $n=1$
and generally improves by increasing the interaction strength. At lower $T/t$,
we take advantage of numerical resummations, such as Euler and Wynn
transformations~\cite{Tang2013}, to obtain an estimate. The NLCE provides a fast
tool, which, given the value of $U/t$, generates results on a dense temperature
and chemical potential grid in a single run.

\subsection{Local density approximation.}  The local density approximation,
which has been previously shown to agree well with \textit{ab initio} DQMC
simulations of the trapped Hubbard Hamiltonian~\cite{Chiesa2011},  was used to
calculate the trap profiles of the different thermodynamic quantities.  The
spin structure factor $S_{\bv{Q}}$ is obtained from the trap profile of the
spin structure factor per lattice site as 
\begin{equation*}
  S_{\bv{Q}} = \frac{1}{Na^{3}} \int s_{\bv{\pi}} \,\mathrm{d}^{3} r. 
\end{equation*}

For the numerical calculations we set $T$ and $\mu_{0}$; local values of $U/t$,
$T/t$, and the local chemical potential $\mu/t$ are calculated using the known
trap potential. The local values of the thermodynamic quantities are then
obtained by interpolation from NLCE and DQMC results for a homogeneous system
calculated in a $(U/t, T /t, \mu/t)$ grid.  Radial profiles for the local value
of  $U/t$, $T/t$, and $\mu/t$ along a body diagonal of the lattice were used
and spherical symmetry assumed.

\subsection{Entropy} In Fig.~4 of the paper we compare the experimental results
at various $U_{0}/t_{0}$ with calculations at constant $T$.   Since ultracold
atoms are isolated systems, a constant value of the overall entropy per particle
$S/(N k_{\mathrm{B}})$ may be more appropriate.  We find that over the range
$10<U_{0}/t_{0}<15$, where AFM correlations are largest,  $S/(N
k_{\mathrm{B}})$ does not vary significantly with $U_{0}/t_{0}$, at constant
$T$ (Extended Data Fig. 6).  This implies that we do not expect significant
adiabatic cooling for stronger interactions~\cite{Werner2005,Paiva2011}, and
thus the curves at constant $T$ are suitable to describe the experimental data.

\newpage
\onecolumngrid

\ExtFigOne
\ExtFigTwo
\ExtFigThree 

\ExtFigFour

\ExtFigFive

\ExtFigSix

\twocolumngrid

\end{document}